\documentclass[reprint,amsmath,amssymb,aps,superscriptaddress,twocolumn,float]{revtex4-1}
\usepackage{graphicx,graphics}
\usepackage{dcolumn}
\usepackage{bm}
\usepackage{amssymb}
\usepackage{amsfonts}
\usepackage{float}
\usepackage{xcolor}
\begin{document}
	
\title{Tunable spin Hall and spin Nernst effects in Dirac line-node semimetals XCuYAs (X=Zr, Hf; Y=Si, Ge)}

\author{Babu Baijnath Prasad}
\affiliation{Department of Physics and Center for Theoretical Physics, National Taiwan University, Taipei 10617, Taiwan\looseness=-1}
\affiliation{Nano Science and Technology Program, Taiwan International Graduate Program, Academia Sinica, Taipei 11529, Taiwan\looseness=-1}

\author{Guang-Yu Guo}
\email{gyguo@phys.ntu.edu.tw}
\affiliation{Department of Physics and Center for Theoretical Physics, National Taiwan University, Taipei 10617, Taiwan\looseness=-1}
\affiliation{Physics Division, National Center for Theoretical Sciences, Hsinchu 30013, Taiwan\looseness=-1}

\date{\today}

\begin{abstract}
	
The quaternary arsenide compounds XCuYAs (X=Zr, Hf; Y= Si, Ge) belong to the vast family of 
the 1111-type quaternary compounds, which possess outstanding physical properties 
ranging from $p$-type transparent semiconductors to high-temperature Fe-based superconductors.
In this paper, we study the electronic structure topology, spin Hall effect (SHE) and spin Nernst effect (SNE) in these
compounds based on density functional theory calculations.
First we find that the four considered compounds are Dirac semimetals with the nonsymmorphic symmetry-protected Dirac 
line nodes along the Brillouin zone boundary $A$-$M$ and $X$-$R$ and low density of states (DOS) near the Fermi level ($E_F$).
Second, the intrinsic SHE and SNE in some of these considered compounds are found to be large.  
In particular, the calculated spin Hall conductivity (SHC) of HfCuGeAs is as large as -514 ($\hbar$/e)(S/cm).
The spin Nernst conductivity (SNC) of HfCuGeAs at room temperature is also large, being -0.73 ($\hbar$/e)(A/m-K).
Moreover, both the magnitude and sign of the SHC and SNC in these compounds can be manipulated 
by varying either the applied electric field direction or spin current direction. The SHE and SNE 
in these compounds can also be enhanced by tuning the Fermi level via chemical doping or electric gating.
Finally, a detailed analysis of the band-decomposed and $k$-resolved spin Berry curvatures reveals  
that these large SHC and SNC as well as their notable tunabilities originate largely from
the presence of a large number of spin-orbit coupling-gapped Dirac points near the Fermi level as well as
the gapless Dirac line-nodes, which give rise to large spin Berry curvatures.
Our findings thus suggest that the four XCuYAs compounds not only provide a valuable platform
for exploring the interplay between SHE, SNE and band topology but also have promising applications 
in spintronics and spin caloritronics. 
	
\end{abstract}

\maketitle

\section{INTRODUCTION}

In recent years, spintronics has emerged as a major field of research in condensed matter physics due to its promising applications 
in energy-efficient data storage and energy harvesting \cite{Puebla2020}.
The key issues in spintronics are the generation, detection and manipulation of spin current.
Indeed, advanced techniques such as injection of spin current from ferromagnets to nonmagnetic materials,
and spin pumping \cite{Switkes1999} have been recently developed that make the generation of spin current become feasible.
On the other hand, the intriguing spin Hall effect (SHE), first predicted by Dyakonov and Perel in 1971 \cite{Dyakonov1971}, 
would provide a pure spin current and does not require a magnetic material or an applied magnetic field.
In a nonmagnetic material under an electric field, due to the influence of the spin-orbit coupling (SOC), 
conduction electrons with opposite spins would acquire opposite transverse velocities, 
thus giving rise to a pure transverse spin current \cite{Murakami2003}.
This pure spin current, generated without an applied magnetic field nor a magnetic material  
is useful for the development of low-power-consumption nanoscale spintronic devices \cite{Liu2012}.
Similarly, in a nonmagnetic material,  one could also get a pure spin current when a temperature gradient ($\nabla T$) 
instead of an electric field $E$ is applied, and this is known as the spin Nernst effect (SNE) \cite{Cheng2008}. 
The SNE thus makes thermally driven spin currents possible and leads to a new field known as the spin caloritronics \cite{Bauer2012}.
Pt \cite{Meyer2017} and tungsten \cite{Sheng2017} are the few well-known examples in which the SNE as well as SHE
have been widely exploited in spintronic research. Clearly, other materials that exhibit large SHE and SNE would 
have wide applications for spintronic and spin caloritronic devices.

Recent theoretical studies indicated that 27$\sim$30 \% of all non-magnetic crystalline materials 
in Inorganic Crystal Structure Database (ICSD) are topological, with roughly 12\% topological insulators 
and 15$\sim$18 \% topological semimetals \cite{Tang2019a,Zhang2019,Vergniory2019,Tang2019b,Hurtubise2019}.
Furthermore, recent theoretical studies on the SHE in Weyl semimetal TaAs~\cite{Sun2016} 
and metallic rutile oxide Dirac semimetals~\cite{Sun2017} revealed that gapped topological nodal points and lines
could lead to a very large spin Hall conductivity (SHC). Also, it was recently reported that the predicted large
SHE and SNE in tetragonal Dirac semimetals ZrXY (X = Si, Ge; Y = S, Se, Te) result from the presence of many slightly gapped
Dirac nodal points near the Fermi level.~\cite{Yen2020}
The four quaternary arsenide XCuYAs (X=Zr, Hf; Y= Si, Ge) compounds considered here have
the same tetragonal symmetry (space group \textit{P}4/\textit{nmm}) as the ZrXY compounds.
As for the ZrXY compounds, they belong to the structure type of the vast family of the so-called 1111-like quaternary phases,
which exhibit outstanding physical properties from $p$-type transparent semiconductors to high-temperature Fe-based superconductors. 
Nevertheless, only few studies on these XCuYAs compounds have been carried out so far,
and these previous studies were focussed mainly on the structural, elastic and electronic properties as well as chemical bonding 
and stability of these quaternary arsenides.~\cite{Johnson1974,Blanchard2010,Baergen2011,Bannikov2012}
Also, the SHE and SNE in these compounds have not been studied yet.  
        
In this work, therefore, we systematically study the electronic structure, SHE and SNE in the XCuYAs (X=Zr, Hf; Y= Si, Ge) compounds
by performing \textit{ab initio} density functional theory (DFT) calculations. The rest of this paper is organized as follows.
In Sec. II, we introduce the crystal structure of the compounds,  briefly describe the Berry phase formalism
for calculating the intrinsic spin Hall and Nernst conductivities as well as the {\it ab initio} computational details.
The main results are presented in Sec. III, which consists of four subsections. In this section, we first examine 
the calculated relativistic band structures and density of states of the four considered compounds in Sec. III A,
which reveal the symmetry-protected Dirac line-nodes as well as a large number of SOC-gapped Dirac nodal points.
We then report the calculated SHC and SNC, and also compare them with that in other well-known materials in Sec. III B and III C, respectively.
In Sec. III D, in order to understand the origins of the large SHC and SNC in some of the considered compounds,
we present the band-decomposed and $k$-resolved spin Berry curvatures for the energy bands near the Fermi level
along the high symmetry lines in the Brillouin zone.
Finally, we give a summary of the conclusions drawn from this work in Sec. IV.

\begin{figure}[htbp] \centering 
\includegraphics[width=8.6cm]{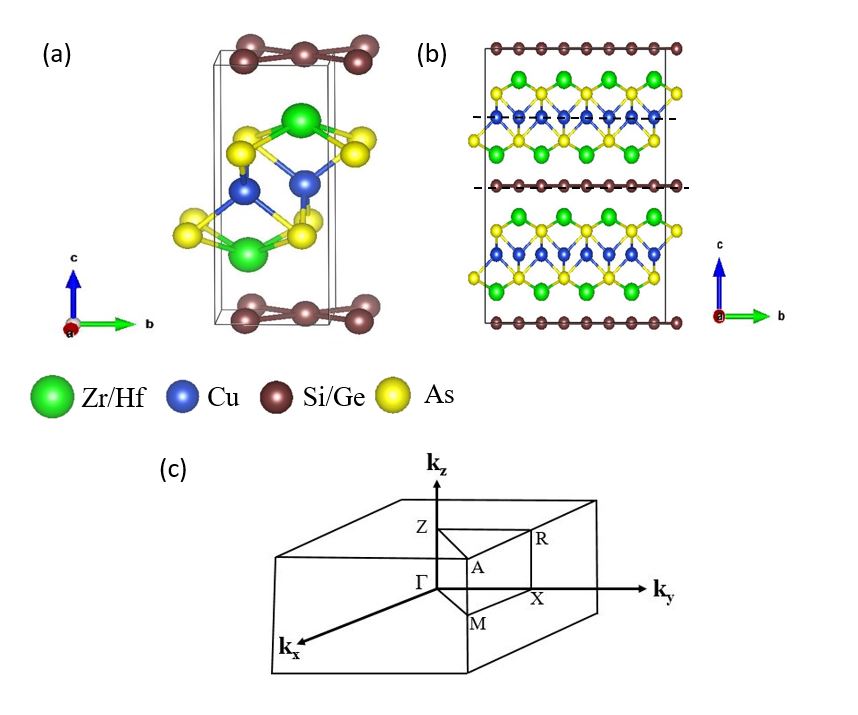}
\caption{(a) Crystal structure of the XCuYAs (X=Zr, Hf; Y= Si, Ge) family, (b) illustration of its nonsymmorphic glide mirror symmetry,
depicting that the Cu and Si/Ge layers serve as the glide mirror planes as indicated by the black dashed lines, and (c) the corresponding
tetragonal Brillouin zone (BZ).}
\label{fig:XCuYAs}
\end{figure}

\section{THEORY AND COMPUTATIONAL DETAILS}

All the four XCuYAs (X=Zr, Hf; Y= Si, Ge) compounds belong to the tetragonal 1111-like quaternary arsenide ZrCuSiAs 
family with space group \textit{P}4/\textit{nmm}. Figure \ref{fig:XCuYAs}(a) shows their crystal structure with two formula units per unit cell.
Table I lists the experimental lattice constants of all the four considered compounds used in the present {\it ab initio} study.
The available experimentally determined atomic positions are also used except the atomic coordinates of HfCuGeAs 
which are determined theoretically while keeping the lattice constants fixed to the experimental values.

Our self-consistent electronic structure calculations as well as the theoretical determination
of the atomic coordinates of HfCuGeAs are based on the DFT with the generalized gradient approximation (GGA) in the form of
Perdew-Burke-Ernzerhof (PBE) \cite{Perdew1996}. 
The accurate projector augmented-wave method \cite{Blochl1994}, 
as implemented out in the Vienna Ab Initio Simulation Package (VASP) \cite{Kresse1993,Kresse1996}, is used.
The valence electronic configurations of Zr, Hf, Cu, Si, Ge and As taken into account in the present study are
$4s^24p^64d^25s^2$,  $5d^26s^2$, $3d^{10}4s^1$, $3s^23p^2$, $3d^{10}4s^24p^2$ and $4s^24p^3$, respectively.
The plane-wave basis set with a large energy cutoff of 450 eV is used.
In the self-consistent band structure calculations, a $\Gamma$-centered $k$-point mesh of 10 $\times$ 10 $\times$ 4 
is used in the BZ integration by the tetrahedron method \cite{Jepson1971,Termmerman1989}.
However, a denser $k$-point grid of 20 $\times$ 20 $\times$ 8 is used for the density of states (DOS) calculations.
      
The spin Hall conductivity (SHC) and spin Nernst conductivity (SNC) are calculated within the elegant Berry-phase formalism
\cite{Guo2008,Xiao2010}.
Within this formalism, the spin Hall conductivity ($\sigma_{ij}^s=J_i^s/E_j$) is simply given as a BZ integration of the
spin Berry curvature for all the occupied bands below the Fermi level ($E_{F}$),
\begin{equation}
\label{eq:1}
\begin{aligned}
\sigma_{ij}^{s}=-e\sum_{n}\int_{BZ}\frac{d\textbf{k}}{(2\pi)^3}f_{\textbf{k}n}\Omega_{ij}^{n,s}(\textbf{k}),
\end{aligned}
\end{equation}

\begin{equation}
\label{eq:2} 
\begin{aligned}
\Omega_{ij}^{n,s}{\textbf{(k)}}=-\sum_{n'\neq n}\frac{2Im\left[\langle\textbf{k}n|\{\tau_{s},v_{i}\}/4|\textbf{k}n'\rangle\langle\textbf{k}n'|v_{j}|\textbf{k}n\rangle \right] }{(\epsilon_{\textbf{k}n}-\epsilon_{\textbf{k}n'})^2},
\end{aligned}
\end{equation}  
where $J_i^s$ is the $i$th component of the spin current density $J^s$, $E_j$ is the $j$th component of the electric field,
$f_{\textbf{k}n}$ is the Fermi distribution function and $\Omega_{ij}^{n,s}{\textbf{(k)}}$ is the spin Berry curvature for the $n^{th}$ band
at $\textbf{k}$, respectively with $i,j\in (x,y,z)$ and $i\neq j$.
Here, $s$ denotes the spin direction. Also, $\tau_s$ and $v_i$ correspond to the Pauli matrix and the velocity operator, respectively
\cite{Guo2005}.
Similarly, the spin Nernst conductivity ($\alpha_{ij}^s=-{J_{i}^{s}}/{\nabla_{j}T}$) can be written as \cite{Xiao2010,Guo2017},
\begin{equation}
\label{eq:3} 
\begin{aligned}
\alpha_{ij}^{s}&= \frac{1}{T}\sum_{n}\int_{BZ}\frac{d\textbf{k}}{(2\pi)^3}\Omega_{ij}^{n,s} (\textbf{k})\\
&\times [(\epsilon_{\textbf{k}n}-\mu)f_{\textbf{k}n}+k_{B}T\ln (1+e^{-\beta(\epsilon_{\textbf{k}n}-\mu)})],
\end{aligned}
\end{equation}
where $\mu$ and $k_B$ are the chemical potential and Boltzmann constant, respectively.

\begin{table*}
\caption{Experimental lattice constants ($a$, $c$), calculated density of states at the Fermi level [$N(E_{F})$] (states/eV/f.u.),
spin Hall conductivity ($\sigma_{xy}^{z}$, $\sigma_{xz}^{y}$ and $\sigma_{zx}^{y}$),
and spin Nernst conductivity ($\alpha_{xy}^{z}$, $\alpha_{xz}^{y}$ and $\alpha_{zx}^{y}$) at temperature $T=300$ K.
Previous results for Dirac semimetals ZrGeTe, noncollinear antiferromagnetic Mn$_{3}$Ge and heavy metal Pt are also listed for comparison.
Note that the unit of the spin Hall conductivity (spin Nernst conductivity) is ($\hbar$/e)(S/cm) [($\hbar$/e)(A/m-K)].}
\begin{ruledtabular}
\begin{tabular}{c c c c c c c c c c}
System &$a$&$c$&$N(E_{F}$) & $\sigma_{xy}^{z}$ & $\sigma_{xz}^{y}$ & $\sigma_{zx}^{y}$ & $\alpha_{xy}^{z}$ & $\alpha_{xz}^{y}$ & $\alpha_{zx}^{y}$ \\
& (\AA ) & (\AA ) &             &             &             &             &              &           &            \\ \hline
ZrCuSiAs & 3.6736$^a$ & 9.5712$^a$ & 0.67 & -45 & -100 & -175 & 0.20 & -0.17 & -0.22 \\
ZrCuGeAs & 3.7155$^b$ & 9.5644$^b$ & 0.56 & -112 & -128 & -193 & 0.16 & -0.32 & -0.54 \\
HfCuSiAs & 3.6340$^a$ & 9.6010$^a$ & 0.65 & -168 & -279 & -400 & 0.46 & -0.45 & -0.49 \\
HfCuGeAs & 3.6935$^b$ & 9.5557$^b$ & 0.57 & -207 & -346 & -514 & 0.53 & -0.52 & -0.73 \\
ZrGeTe$^c$ & -- & -- & 1.12 & 136 & -262 & -551 & 0.53 & 0.48 & 1.17 \\
Mn$_{3}$Ge$^d$ & -- & -- & 2.37 & 56 & -- & -- & 0.14 & -- & -- \\
Pt & -- & -- & 1.75 & 2139$^e$ & -- & -- & -1.09 (-0.91)$^d$,-1.57$^{f}$ & -- & -- \\
\end{tabular}
\end{ruledtabular}
{$^{a}$X-ray diffraction experiment \cite{Johnson1974};
 $^{b}$X-ray diffraction experiment \cite{Baergen2011};
 $^{c}$\textit{Ab initio} calculation \cite{Yen2020};
 $^{d}$\textit{Ab initio} calculation \cite{Guo2017};
 $^{e}$\textit{Ab initio} calculation \cite{Guo2008};
 $^{f}$ Experiment at 255 K \cite{Meyer2017}.}
\label{table:1}
\end{table*}

In the spin Hall and spin Nernst conductivities calculations, the velocity ($\langle\textbf{k}n'|v_{i}|\textbf{k}n\rangle$)
and spin velocity ($\langle\textbf{k}n|\{\tau_{s},v_{i}\}/4]|\textbf{k}n'\rangle$) matrix elements are obtained from the self-consistent
relativistic band structures within the PAW formalism \cite{Adolph2001}. A homemade program \cite{Guo2017,Guo2014} is used for the band summation
and the Brillouin zone (BZ) integration with the tetrahedron method.~\cite{Termmerman1989}
Very fine grids of 45384 (for ZrCuSiAs and HfCuGeAs), 43493 for HfCuSiAs and 106272 for ZrCuGeAs $k$ points in the irreducible BZ wedge
are used and correspond to the division of $\Gamma X$ line into $n_d=60$ (For ZrCuSiAs, HfCuSiAs and HfCuGeAs)
and $n_d=80$ (for ZrCuGeAs) intervals, respectively.
Test calculations using several different sets of $k$-point meshes show that such calculated values of the SHC and SNC 
for all the four considered compounds converge within a few percent.

\section{RESULTS AND DISCUSSION}

\subsection{Electronic structure}

\begin{figure}[htbp] \centering
\includegraphics[width=8.85cm]{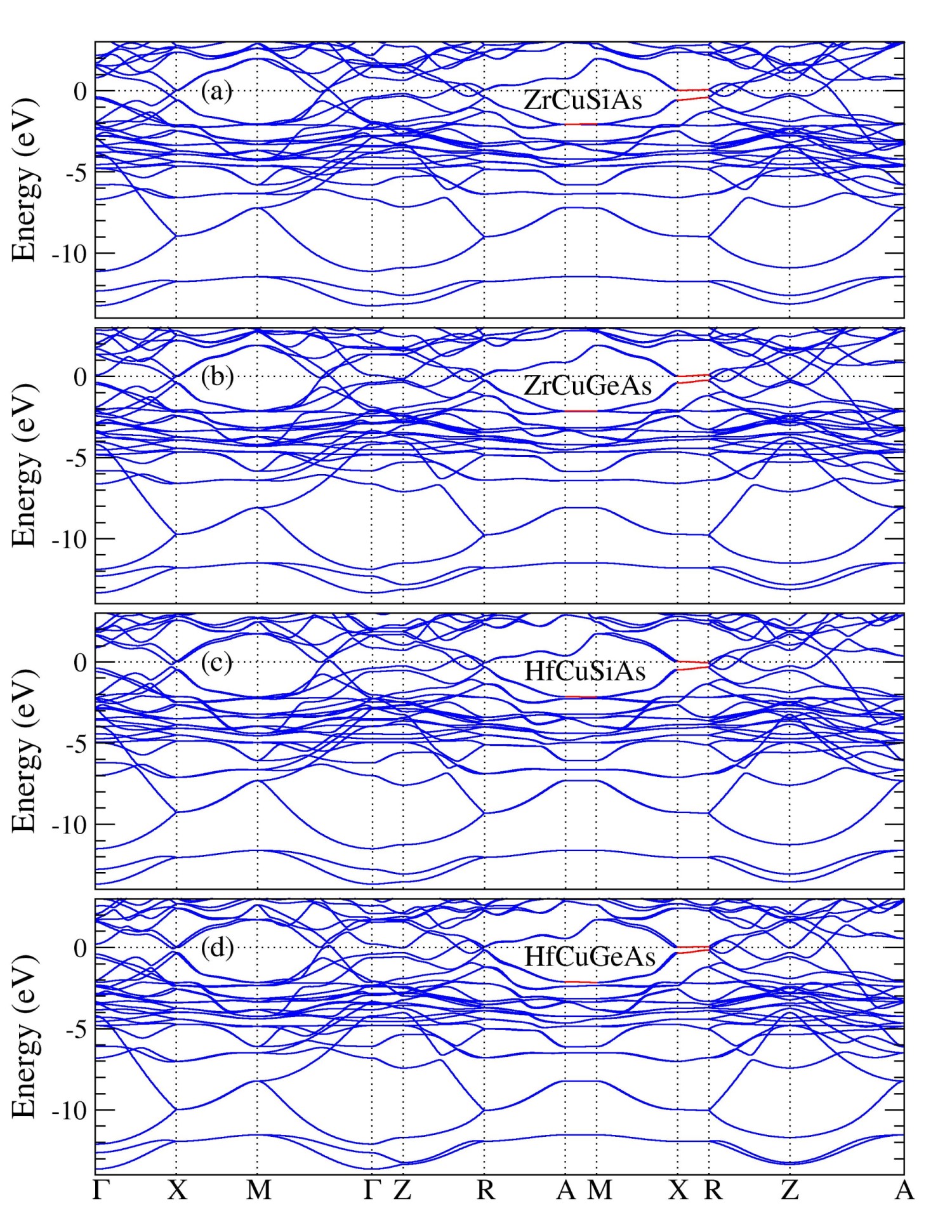}
\caption{Relativistic band structures of (a) ZrCuSiAs, (b) ZrCuGeAs, (c) HfCuSiAs, and (d) HfCuGeAs. 
In (a), (b), (c), and (d), the Fermi level is at the zero energy. The red curves represent the Dirac line-nodes.}
\label{fig:XCuYAs-band}
\end{figure}

\begin{figure}[htbp] \centering
\includegraphics[width=8.85cm]{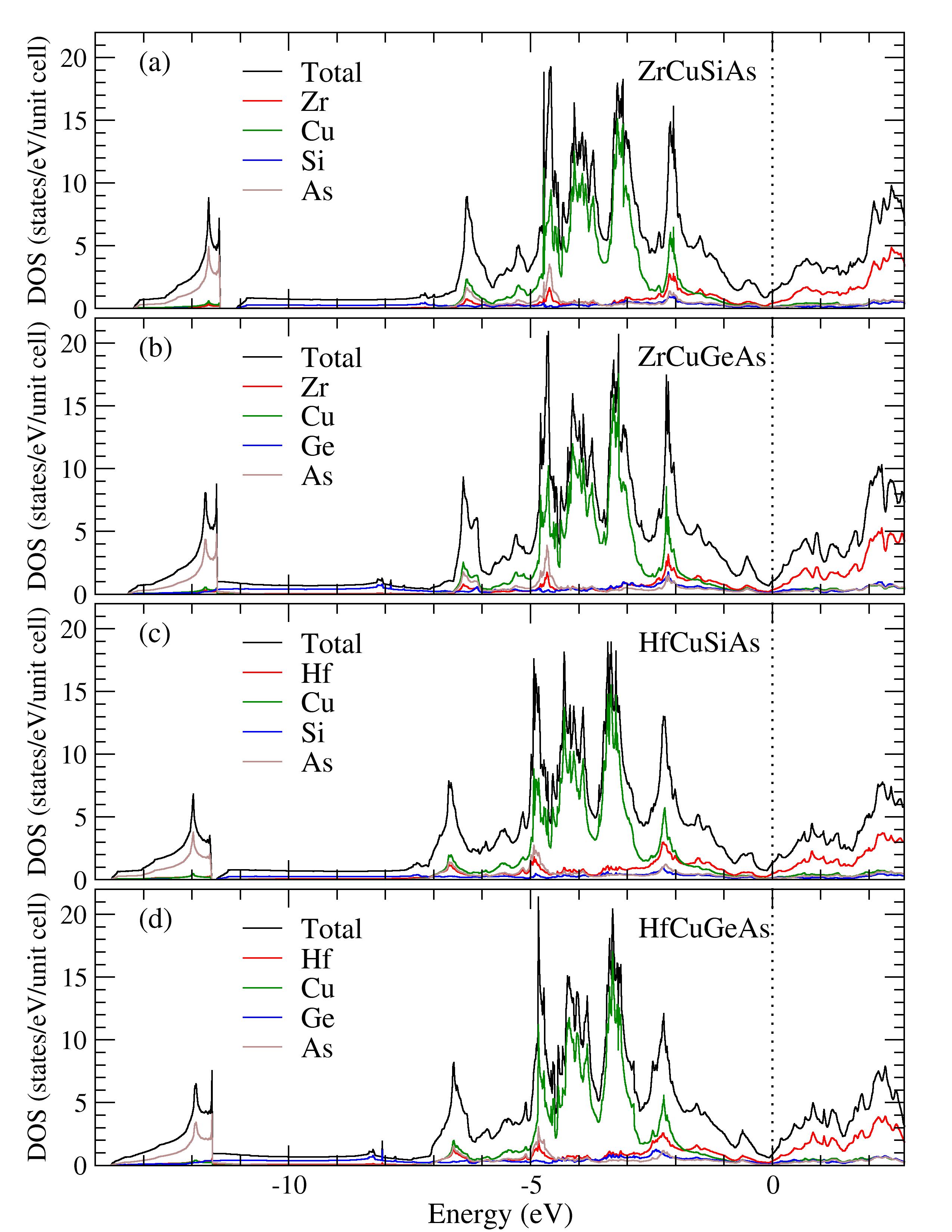}
\caption{Total and atom-decomposed density of states (DOS) of (a) ZrCuSiAs, (b) ZrCuGeAs, (c) HfCuSiAs, and (d) HfCuGeAs. 
In (a), (b), (c), and (d), the Fermi level is at the zero energy.}
\label{fig:XCuYAs-dos}
\end{figure}

Our calculations show that all the four XCuYAs (X=Zr, Hf; Y= Si, Ge) compounds are nonmagnetic. 
Their calculated relativistic band structures and density of states (DOS) are displayed, respectively, 
in Figs. \ref{fig:XCuYAs-band} and \ref{fig:XCuYAs-dos}. 
Figure 3 shows that the lowest two valence bands of the compounds (see Fig. 2) are derived from As $s$-orbitals.
Above these bands are the dispersive Si (Ge) $sp$-dominated valence bands in ZrCuSiAs and HfCuSiAs (ZrCuGeAs and HfCuGeAs)
(see Figs. 2 and 3). The valence bands in between -2.0 to -6.0 eV in ZrCuSiAs and HfCuSiAs (ZrCuGeAs and HfCuGeAs)
are mainly contributed from the Cu $d$ orbitals
together with small contributions from the Zr (Hf) $d$ orbital, Si (Ge) $p$ orbital and As $p$ orbital. 
On the other hand,
the valence bands and conduction bands in the vicinity of the Fermi level ($E_F$) (Fig. 2) are mainly
contributed from either the Zr $d$ orbital for ZrCuSiAs and ZrCuGeAs or the  Hf $d$ orbital for HfCuSiAs and HfCuGeAs, 
as can be seen from the atom-decomposed DOS spectra in Fig. 3.
Figure 2 also show that all the four considered compounds are semimetals with few bands crossing $E_F$ 
and thus have low DOS values near $E_F$ (see Fig. 3).

Importantly, Fig. \ref{fig:XCuYAs-band} reveals that all the four considered XCuYAs compounds possess a number of slightly gapped 
Dirac points near $E_{F}$ along the $\Gamma$-$X$, $M$-$\Gamma$, $Z$-$R$, and $Z$-$A$ lines in the BZ
(see also Figs. \ref{fig:ZrCuSiAs}(a), \ref{fig:ZrCuGeAs}(a), \ref{fig:HfCuSiAs}(a), and \ref{fig:HfCuGeAs}(a) below).
In the calculated scalar-relativistic band structures (i.e., without SOC) (not shown here), these Dirac points are gapless.
In the presence of SOC, however, the number of irreducible representations of the $C_{2v}$ point group is reduced to one 
and consequently, these Dirac points are gapped out.
As will be shown in Sec. III D, these gapped Dirac points exhibit large spin Berry curvatutes and thus
make significant contributions to the SHC and SNC [see Eqs. (1) and (3)].
Interestingly, we also find that there are Dirac line-nodes along the
$A$-$M$ and $X$-$R$ on the BZ boundary (see the red curves in Fig. 2), similar to that of the ZrXY family \cite{Yen2020}.
These Dirac line-nodes are protected by the nonsymmorphic symmetry and thus are not gapped out in the presence of SOC.
The Dirac line-nodes along the $A$-$M$ are 2.0 eV below $E_F$ and thus would not significantly
influence the transport properties such as SHC and SNC. In contrast, the Dirac line-nodes along the $X$-$R$
are close to the Fermi level and thus will have significant effects on the transport properties of these compounds.

\subsection{Spin Hall effect}

The SHC of a solid is a third-rank tensor ($\sigma_{ij}^{s}$; $s,i,j=x,y,z$).
Nevertheless, only a few of elements of the SHC tensor are nonvanishing because of the symmetry constraints.
For the nonmagnetic XCuYAs (X=Zr, Hf; Y= Si, Ge) compounds having space group \textit{P}4/\textit{nmm}, 
as revealed by a recent symmetry analysis \cite{Seemann2015}, there are only five non-zero tensor elements. 
Also note that because of the tetragonal symmetry of these compounds, the $x$ and $y$ axes are equivalent 
and this results in $\sigma_{yz}^{x} = -\sigma_{xz}^{y}$ and $\sigma_{zx}^{y} = -\sigma_{zy}^{x}$.
Therefore, we end up with only three independent nonzero elements, namely, $\sigma_{xy}^{z}$, $\sigma_{xz}^{y}$ and $\sigma_{zx}^{y}$
where superscripts $x$, $y$ and $z$ denote, respectively, the spin direction of the spin current 
along the [100], [010] and [001] directions, as listed in Table II.
The calculated values of all the nonzero SHC tensor elements for the XCuYAs family are listed in Table I.
In Table I, $\sigma_{xy}^{z}$, $\sigma_{xz}^{y}$ and $\sigma_{zx}^{y}$ of a recently studied Dirac semimetal ZrGeTe \cite{Yen2020},
as well as $\sigma_{xy}^{z}$ of noncollinear antiferromagnet Mn$_3$Ge \cite{Guo2017} and platinum metal \cite{Meyer2017,Guo2008}, 
are also listed for comparison.

\begin{table}
\caption{Symmetry-imposed tensor forms of the spin Hall conductivity (SHC) tensor for space group \textit{P}4/\textit{nmm}\cite{Seemann2015}.
The forms of the spin Nernst conductivity tensor are identical to the corresponding forms of the SHC tensor.}
\begin{tabular}{c c c}
\hline\hline
& SHC & \\ 
\underline{$\sigma$}$^x$ & \underline{$\sigma$}$^y$ & \underline{$\sigma$}$^z$ \\ 
\hline
\\
$\left(\begin{array}{ccc} 0 & 0 & 0\\0 & 0 & -\sigma_{xz}^{y}\\0 & -\sigma_{zx}^{y}&0\end{array}\right)$ &
$\left(\begin{array}{ccc} 0 & 0 & \sigma_{xz}^{y}\\0 & 0 & 0\\\sigma_{zx}^{y} & 0&0\end{array}\right)$ &
$\left(\begin{array}{ccc} 0 & \sigma_{xy}^{z} & 0\\-\sigma_{xy}^{z} & 0 & 0\\0 & 0&0\end{array}\right)$ \\ \\
\hline\hline
\end{tabular}
\label{table:2}
\end{table}

\begin{figure}[htbp] \centering
\includegraphics[width=8.0cm]{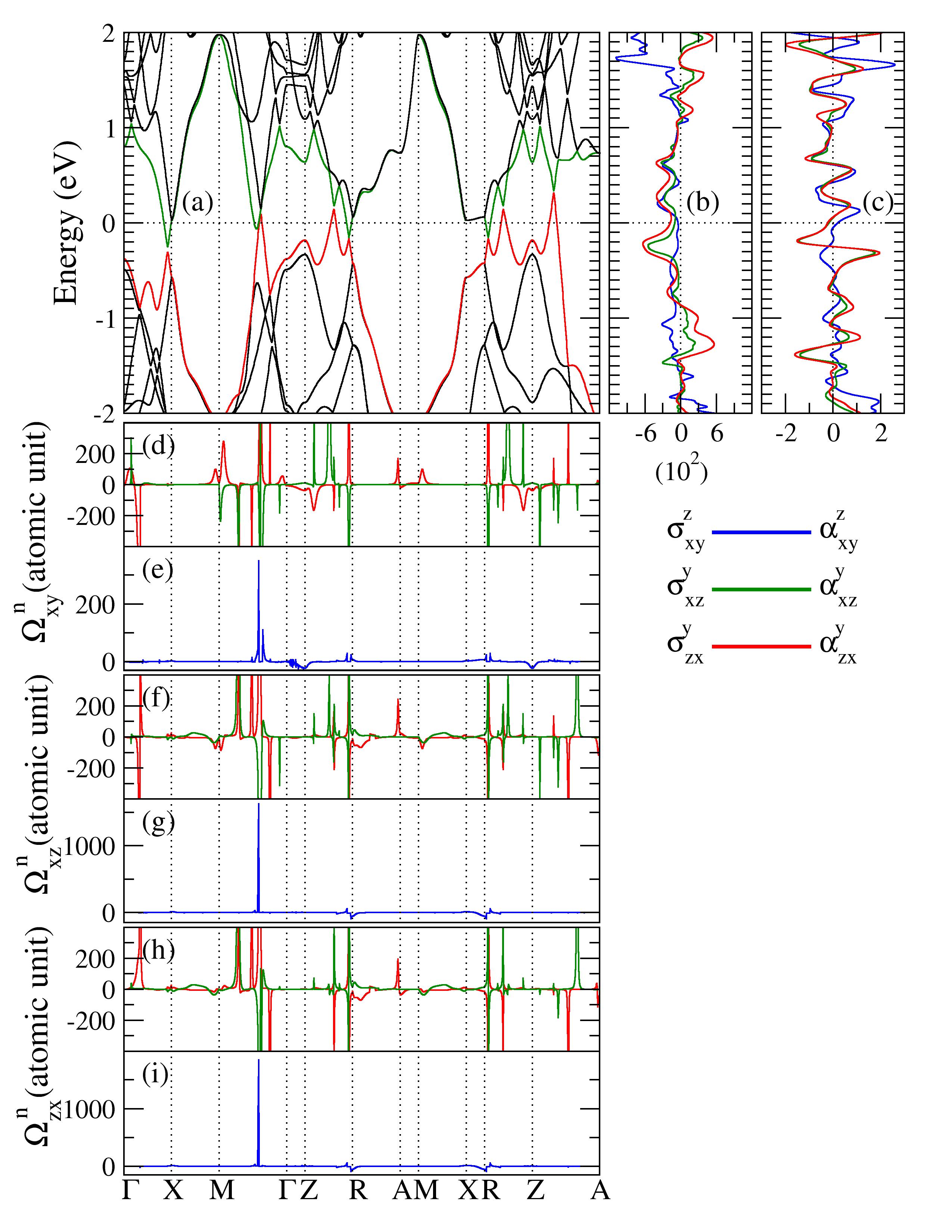}
\caption{ZrCuSiAs. (a) Relativistic band strucuture,
(b) spin Hall conductivity (SHC; $\sigma_{xy}^{z}$, $\sigma_{xz}^{y}$ and $\sigma_{zx}^{y}$) as a function of energy,
(c) spin Nernst conductivity (SNC; $\alpha_{xy}^{z}$, $\alpha_{xz}^{y}$ and $\alpha_{zx}^{y}$) at $T=300$ K as a function of energy,
(d), (f) and (h) band-decomposed spin Berry curvatures ($\Omega^{n}$),
as well as (e), (g) and (i) total spin Berry curvatures ($\Omega$) along the high symmetry lines in the Brillouin zone. 
In (a), (b) and (c), the Fermi level is at the zero energy,
and the unit of the SHC (SNC) is ($\hbar$/e)(S/cm) [($\hbar$/e)(A/m-K)].
Note that in (a), (d), (f) and (h), the same color curves correspond to the same bands.}
\label{fig:ZrCuSiAs}
\end{figure}

\begin{figure}[htbp] \centering
\includegraphics[width=8.0cm]{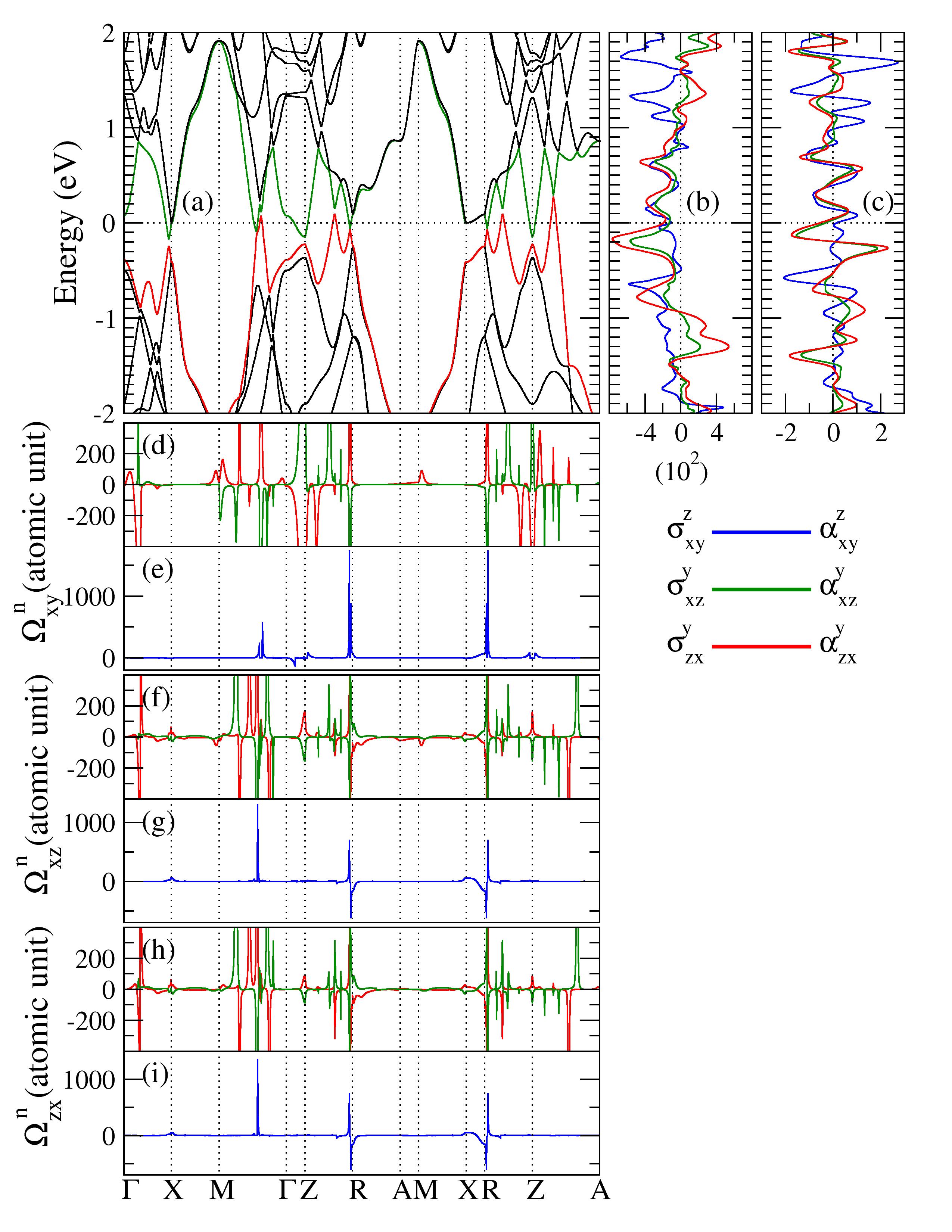}
\caption{ZrCuGeAs. (a) Relativistic band strucuture,
(b) spin Hall conductivity (SHC; $\sigma_{xy}^{z}$, $\sigma_{xz}^{y}$ and $\sigma_{zx}^{y}$) as a function of energy,
(c) spin Nernst conductivity (SNC; $\alpha_{xy}^{z}$, $\alpha_{xz}^{y}$ and $\alpha_{zx}^{y}$) at $T=300$ K as a function of energy,
(d), (f) and (h) band-decomposed spin Berry curvatures ($\Omega^{n}$),
as well as (e), (g) and (i) total spin Berry curvatures ($\Omega$) along the high symmetry lines in the Brillouin zone. 
In (a), (b) and (c), the Fermi level is at the zero energy,
and the unit of the SHC (SNC) is ($\hbar$/e)(S/cm) [($\hbar$/e)(A/m-K)].
Note that in (a), (d), (f) and (h), the same color curves correspond to the same bands.}
\label{fig:ZrCuGeAs}
\end{figure}

\begin{figure}[htbp] \centering
\includegraphics[width=8.0cm]{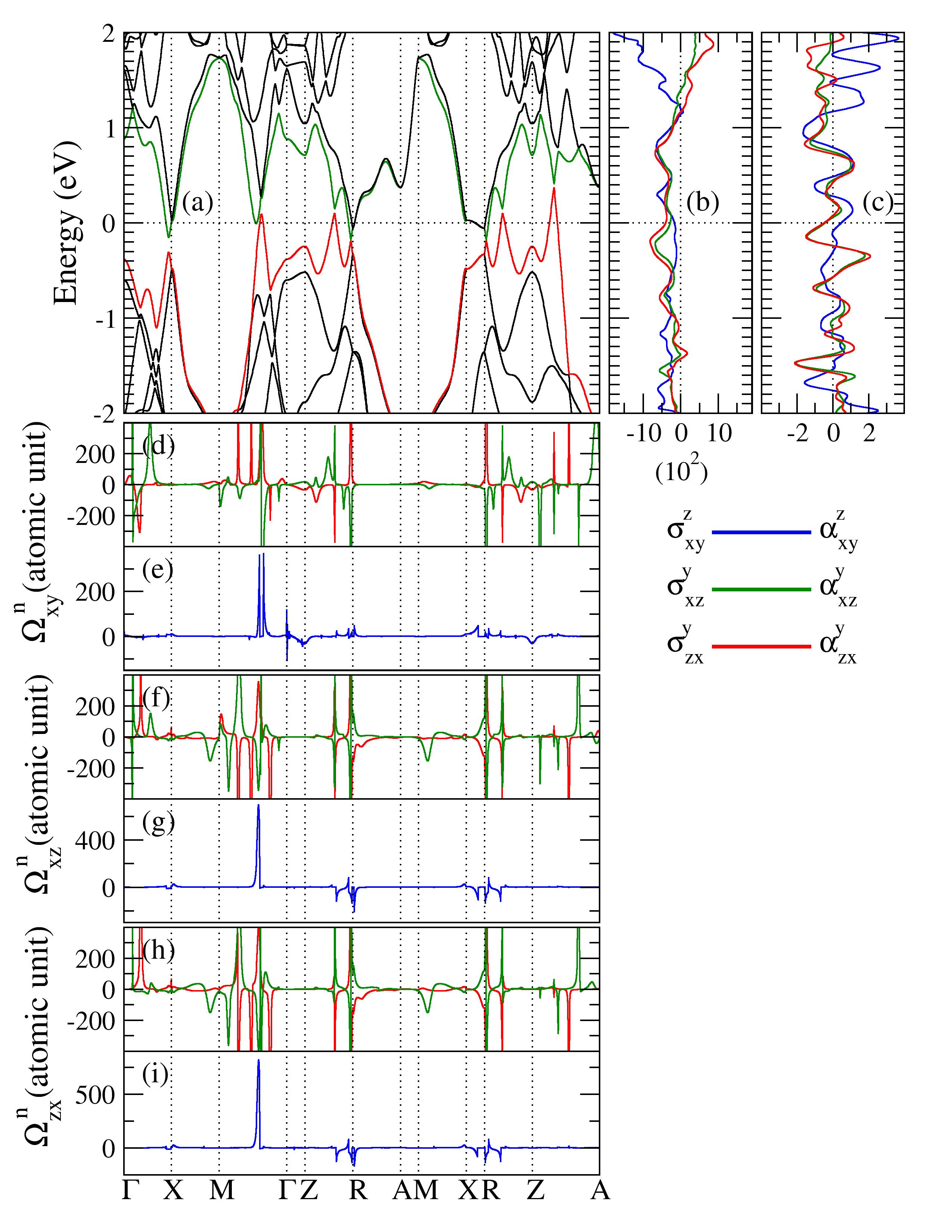}
\caption{HfCuSiAs. (a) Relativistic band strucuture,
(b) spin Hall conductivity (SHC; $\sigma_{xy}^{z}$, $\sigma_{xz}^{y}$ and $\sigma_{zx}^{y}$) as a function of energy,
(c) spin Nernst conductivity (SNC; $\alpha_{xy}^{z}$, $\alpha_{xz}^{y}$ and $\alpha_{zx}^{y}$) at $T=300$ K as a function of energy,
(d), (f) and (h) band-decomposed spin Berry curvatures ($\Omega^{n}$),
as well as (e), (g) and (i) total spin Berry curvatures ($\Omega$) along the high symmetry lines in the Brillouin zone.
In (a), (b) and (c), the Fermi level is at the zero energy,
and the unit of the SHC (SNC) is ($\hbar$/e)(S/cm) [($\hbar$/e)(A/m-K)].
Note that in (a), (d), (f) and (h), the same color curves correspond to the same bands.}
\label{fig:HfCuSiAs}
\end{figure}
 
\begin{figure}[htbp] \centering
\includegraphics[width=8.0cm]{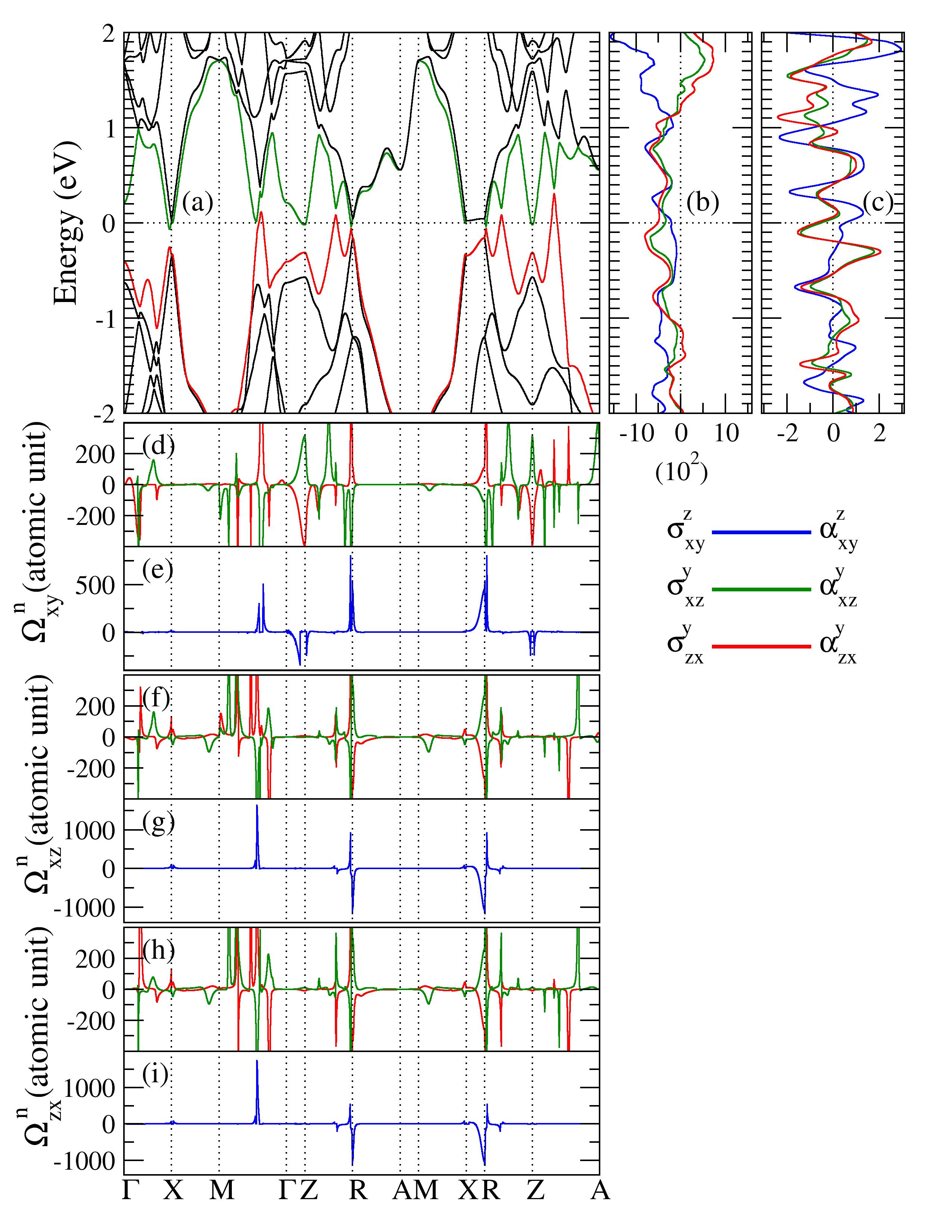}
\caption{HfCuGeAs. (a) Relativistic band strucuture,
(b) spin Hall conductivity (SHC; $\sigma_{xy}^{z}$, $\sigma_{xz}^{y}$ and $\sigma_{zx}^{y}$) as a function of energy,
(c) spin Nernst conductivity (SNC; $\alpha_{xy}^{z}$, $\alpha_{xz}^{y}$ and $\alpha_{zx}^{y}$) at $T=300$ K as a function of energy,
(d), (f) and (h) band-decomposed spin Berry curvatures ($\Omega^{n}$),
as well as (e), (g) and (i) total spin Berry curvatures ($\Omega$) along the high symmetry lines in the Brillouin zone.
In (a), (b) and (c), the Fermi level is at the zero energy,
and the unit of the SHC (SNC) is ($\hbar$/e)(S/cm) [($\hbar$/e)(A/m-K)].
Note that in (a), (d), (f) and (h), the same color curves correspond to the same bands.}
\label{fig:HfCuGeAs}
\end{figure}

In Table I, we list the results for the four XCuYAs compounds in the increasing order of their SOC strength.
Among them, ZrCuSiAs has the weakest SOC strength, whereas HfCuGeAs has the strongest one.
This can be easily understood by a well-known fact that the SOC strength is directly proportional to the fourth power 
of the atomic number of the element.
It is clear from Table I that all the nonzero SHC elements, i.e., $\sigma_{xy}^{z}$, $\sigma_{xz}^{y}$ and $\sigma_{zx}^{y}$ increase
monotonically as we go down from ZrCuSiAs to HfCuGeAs, and this is consistent with the SOC strength among these materials.
The calculated value of $\sigma_{zx}^{y}$ for HfCuGeAs is the largest, being -514 ($\hbar$/e)(S/cm)
and about four times smaller than the SHC ($\sigma_{xy}^{z}$) [2139 ($\hbar$/e)(S/cm)] of platinum metal 
which has the largest intrinsic SHC among the transition metals \cite{Guo2008,Guo2014}. 
Interestingly, this is 9 times larger than that [56 ($\hbar$/e)(S/cm)] of the noncollinear antiferromagnet Mn$_3$Ge \cite{Guo2017} 
and is comparable to the calculated value of $\sigma_{zx}^{y}$ of ZrGeTe which has the largest SOC strength among the ZrXY family \cite{Yen2020}.
Also, $\sigma_{zx}^{y}$ of HfCuSiAs [-400 ($\hbar$/e)(S/cm)] is 1.5 times larger than that [-262 ($\hbar$/e)(S/cm)] of $\sigma_{xz}^{y}$ of
ZrGeTe \cite{Yen2020}.

There is a strong anisotropy in the SHC in the XCuYAs compounds, as displayed in Table I.
For example, the SHC value of ZrCuSiAs can be increased by a factor of 4 by changing the spin polarization 
of the spin current from the $z$ (out-of-plane) direction to either the $x$ or $y$ (in-plane) direction.
Also, for HfCuSiAs and HfCuGeAs, the SHC can be increased by a factor of approximately 2.4 and 2.5, respectively,
when the spin polarization is switched from the $z$ to either the $x$ or $y$ direction.
The calculated SHC for the XCuYAs compounds also depends strongly on the direction of the applied electric field,
as can be seen from Table I.
For all the four compounds, the SHC value for the electric field applied along the [100] direction ($\sigma_{zx}^{y}$)
is approximately 1.5 times larger than that along the [001] direction ($\sigma_{xz}^{y}$; see Table I).

We have also calculated the SHC as a function of Fermi energy ($E_F$) within the rigid-band approximation 
in which only the Fermi energy is varied while keeping the band structure fixed. 
The calculated SHC spectra are presented in Figs. \ref{fig:ZrCuSiAs}(b), \ref{fig:ZrCuGeAs}(b), \ref{fig:HfCuSiAs}(b), and \ref{fig:HfCuGeAs}(b) for ZrCuSiAs, ZrCuGeAs, HfCuSiAs and HfCuGeAs, respectively. 
It is clear that most of the nonzero elements of the SHC tensor have peaks in the vicinity of the Fermi energy,
suggesting that the SHE in these compounds can be optimized by varying $E_{F}$ via either chemical doping or electrical gating.
For example, $\sigma_{xy}^{z}$ of ZrCuGeAs can be increased by a factor of 2 by slightly lowering the Fermi energy by 0.05 eV
[see Fig. \ref{fig:ZrCuGeAs}(b)]. This can be achieved by hole doping of merely 0.03 e/f.u.
Also, $\sigma_{zx}^{y}$ of HfCuGeAs could be enhanced from -514 ($\hbar$/e)(S/cm) to -801 ($\hbar$/e)(S/cm) 
by lowering the Fermi energy to -0.14 eV [see Fig. \ref{fig:HfCuGeAs}(b)], 
and this can be realized via hole doping of 0.07 e/f.u.

\subsection{Spin Nernst effect}

Similar to the SHC for a solid, SNC ($\alpha_{ij}^{s}$; $s,i,j=x,y,z$) is also a third-rank tensor.
Furthermore, the nonzero elements of the SNC tensor are identical to that of the SHC tensor~\cite{Seemann2015}.
Therefore, for the XCuYAs (X=Zr, Hf; Y= Si, Ge) compounds, 
the SNC has only three nonzero independent tensor elements, namely, $\alpha_{xy}^{z}$, $\alpha_{xz}^{y}$ and $\alpha_{zx}^{y}$.
We list the calculated values of these nonzero SNC elements at $T = 300$ K in the XCuYAs family in Table I.
We notice that $\alpha_{zx}^{y}$ of HfCuGeAs is the largest [-0.73 ($\hbar$/e)(A/m-K)] among the XCuYAs family.
Also, it is almost 5 times larger than $\alpha_{xy}^{z}$ [0.14 ($\hbar$/e)(A/m-K)] of the noncollinear antiferromagnet Mn$_3$Ge \cite{Guo2017}.
The $\alpha_{xy}^{z}$ values of HfCuGeAs and ZrGeTe \cite{Yen2020} are equal.
They are comparable to that of Pt metal, being about half the $\alpha_{xy}^{z}$ value of Pt (Table I).
Therefore, we can say that the SNC values for the XCuYAs compounds are significant.

The SNC of the four XCuYAs compounds is also highly anisotropic, similar to the SHC.
For example, the $\alpha_{zx}^{y}$ of ZrCuGeAs is approximately 3.4 times larger than $\alpha_{xy}^{z}$ of the same material (see Table I),
i.e., the SNC value of ZrCuGeAs is enhanced by a factor of 3.4 when the spin direction of the spin current is switched 
from the $z$ to the $x$ or $y$ direction.
Unlike the SHC, the SNC of the XCuYAs family changes sign when the spin direction of the spin current is rotated 
from the out-of-plane ($z$) direction to an in-plane ($x$ or $y$) direction.
For HfCuGeAs, $\alpha_{zx}^{y}$ is 1.4 times larger than $\alpha_{xz}^{y}$ (see Table I).

The SNC at $T = 300$ K has also been calculated as a function of $E_{F}$ and is displayed in Fig. \ref{fig:ZrCuSiAs}(c), \ref{fig:ZrCuGeAs}(c),
\ref{fig:HfCuSiAs}(c) and \ref{fig:HfCuGeAs}(c), respectively, for ZrCuSiAs, ZrCuGeAs, HfCuSiAs and HfCuGeAs.
These SNC spectra will allow one to optimize the SNE in the XCuYAs compounds via manipulation of $E_{F}$. 
For example, Fig. \ref{fig:ZrCuGeAs}(c) indicates that the $\alpha_{xy}^{z}$ of ZrCuGeAs has a prominent peak of 1.02 ($\hbar$/e)(A/m-K)
at 0.08 eV (just slightly above $E_{F}$). This suggests that the SNC $\alpha_{xy}^{z}$ of ZrCuGeAs can be
substantially increased from 0.16 to 1.02 ($\hbar$/e)(A/m-K) via electron doping of merely 0.05 e/f.u. 
Note that this value of $\alpha_{xy}^{z}$ is very close to that of platinum metal [-1.09 ($\hbar$/e)(A/m-K)] \cite{Guo2017},
thus indicating the possible application of these materials in spin caloritronic devices.
The $\alpha_{zx}^{y}$ of HfCuGeAs also has a peak of -1.53 ($\hbar$/e)(A/m-K) at -0.10 eV (slightly below $E_{F}$),
and this could be achieved by hole doping of merely 0.06 e/f.u. [see Fig. \ref{fig:HfCuGeAs}(c)].
The SNC spectra of the XCuYAs family display a stronger dependence on $E_{F}$ than that of the SHC.
For example, $\alpha_{zx}^{y}$ of ZrCuSiAs decreases rapidly as $E_{F}$ decreases and then reaches the negative local maximum of
-1.51 ($\hbar$/e)(A/m-K) at -0.19 eV.
As we decrease $E_{F}$ further, it increases steadily and changes sign at -0.25 eV and then hits the positive local maximum of
1.96 ($\hbar$/e)(A/m-K) at -0.32 eV [see Fig. \ref{fig:ZrCuSiAs}(c)].
For ZrCuGeAs, $\alpha_{zx}^{y}$ decreases as $E_{F}$ decreases and reaches a negative local maximum of -1.74 ($\hbar$/e)(A/m-K) at -0.12 eV.
It starts to increase as $E_{F}$ is lowered further, changes sign at -0.19 eV and then rise to a positive local maximum of
2.30 ($\hbar$/e)(A/m-K) at -0.26 eV [see Fig. \ref{fig:ZrCuGeAs}(c)].
The $\alpha_{xy}^{z}$ for the four XCuYAs compounds shows a similar behavior
[see Figs. \ref{fig:ZrCuSiAs}(c) - \ref{fig:HfCuGeAs}(c)].
All these suggest that electric gating or chemical doping can be used to manipulate the SHE and SNE in the XCuYAs compounds.

\begin{figure}[htbp] \centering
\includegraphics[width=11.5cm]{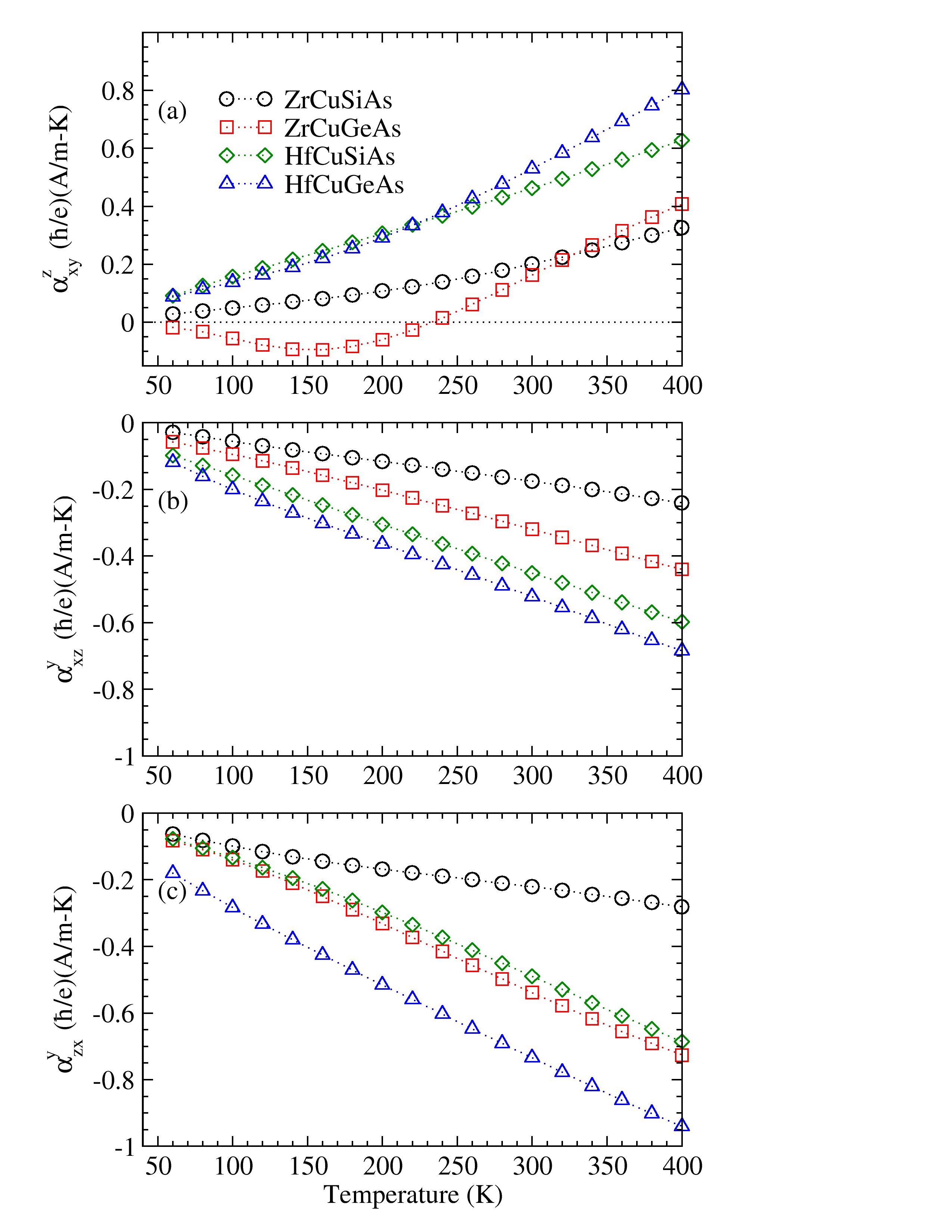}
\caption{Spin Nernst conductivity (a) $\alpha_{xy}^{z}$, (b) $\alpha_{xz}^{y}$, and (c) $\alpha_{zx}^{y}$ as a function 
of temperature $T$ for the four XCuYAs compounds.}
\label{fig:T-scan}
\end{figure}

Temperature ($T$) dependence of the SNC of all the four XCuYAs compounds is also calculated in the present study. 
Figure \ref{fig:T-scan} shows the calculated $T$ dependences of $\alpha_{xy}^{z}$, $\alpha_{xz}^{y}$, and $\alpha_{zx}^{y}$. 
We notice that both $\alpha_{xz}^{y}$ and $\alpha_{zx}^{y}$ for all the four compounds are negative and 
their magnitudes increase monotonically as $T$ increases from $T=60$ K to 400 K [see Figs. \ref{fig:T-scan}(b) and \ref{fig:T-scan}(c)].
Also, HfCuGeAs has the largest magnitude of $\alpha_{xz}^{y}$ and $\alpha_{zx}^{y}$ among the four compounds in the entire considered $T$ range.
The values of $\alpha_{xz}^{y}$ and $\alpha_{zx}^{y}$ for HfCuGeAs hit the negative maximum of
-0.68 and -0.94 ($\hbar$/e)(A/m-K), respectively,  at $T=400$ K (see Fig. \ref{fig:T-scan}).
Interestingly, the $\alpha_{zx}^{y}$ spectra of ZrCuGeAs and HfCuSiAs have almost an identical behavior 
and their calculated values are close in the entire considered temperature range.
In contrast, the $\alpha_{xy}^{z}$ spectra of ZrCuSiAs, HfCuSiAs and HfCuGeAs have a positive value and all of them increase 
monotonically with $T$ [see Fig. \ref{fig:T-scan}(a)].
Interestingly, the values of $\alpha_{xy}^{z}$ for HfCuSiAs and HfCuGeAs are nearly identical for $T \le 260$ K.
However, $\alpha_{xy}^{z}$ of ZrCuGeAs has a negative value at $T=60$ K. It decreases monotonically with $T$ 
and reaches the negative local maximum of -0.095 ($\hbar$/e)(A/m-K) at $T=160$ K.
As temperature $T$ further increases, it starts to increase, then changes sign at $\sim$240 K and finally reaches the maximum value of
0.41 ($\hbar$/e)(A/m-K) at $T=400$ K [see Fig. \ref{fig:T-scan}(a)].
 
\subsection{Spin Berry curvature analysis}

Spin Berry curvature is the key ingredient for calculating spin Hall and spin Nernst conductivities, 
as can be seen from equations (\ref{eq:1}) and (\ref{eq:3}), respectively.
Thus, in order to understand the origin of the SHC and SNC for the XCuYAs compounds, we have calculated total and band-decomposed 
spin Berry curvatures for the energy bands near the Fermi level as a function of $k$-point.
The total spin Berry curvature at ${\bf k}$ is the sum of the spin Berry curvatures of all the occupied bands at ${\bf k}$.
Figs. \ref{fig:ZrCuSiAs}(d, f and h), Figs. \ref{fig:ZrCuGeAs}(d, f and h), Figs. \ref{fig:HfCuSiAs}(d, f and h), and
Figs. \ref{fig:HfCuGeAs}(d, f and h) show these spin Berry curvatures $\Omega_{xy}^{z}$, $\Omega_{xz}^{y}$,
and $\Omega_{zx}^{y}$, respectively, along the high symmetry lines in the BZ.

As mentioned before, in the XCuYAs compounds, a number of SOC-gapped Dirac points are present near the Fermi level 
along the $\Gamma$-$X$, $M$-$\Gamma$, $Z$-$R$ and $Z$-$A$ directions in the BZ 
[see Fig. \ref{fig:XCuYAs-band} as well as Figs. \ref{fig:ZrCuSiAs}(a) - \ref{fig:HfCuGeAs}(a)].
As reported previously in the literature \cite{Guo2008,Yen2020}, when a band-crossing point (e.g., a Dirac point) at ${\bf k}$
is slightly gapped by the SOC, a pair of spin Berry curvature peaks with opposite signs appear in the vicinity of the $k$ point 
on the lower and upper gapped bands, respectively. These phenomena can be clearly seen in
Figs. \ref{fig:ZrCuSiAs}(d, f and h), Figs. \ref{fig:ZrCuGeAs}(d, f and h), Figs. \ref{fig:HfCuSiAs}(d, f and h), and
Figs. \ref{fig:HfCuGeAs}(d, f and h).
Clearly, a large contribution to the SHC may occur when only the lower band is occupied.
If both bands are occupied, the two peaks with opposite signs would cancel each other, 
thus leading to little contribution to the SHC \cite{Yen2020}.

Table I shows that HfCuGeAs has the largest values of the nonzero elements of the SHC tensor. 
Therefore, as an example, let us take a closer look at the energy bands near the Fermi level and their spin Berry curvatures
in HfCuGeAs. First we notice that for total spin Berry curvature $\Omega_{xy}^{z}$, there are sharp positive
peaks at the $R$ point as well as near the middle of the $M$-$\Gamma$ line [see the blue curve in Fig. \ref{fig:HfCuGeAs}(e)].
In the meantime, we can see from Fig. \ref{fig:HfCuGeAs}(a) that there are gapped Dirac points 
at these $k$-points, which generate two pairs of sharp  $\Omega_{xy}^{n}$ peaks with opposite signs 
at the same $k$-points [see the red and green curves in Fig. \ref{fig:HfCuGeAs}(d)]. 
Since only the lower band (red curve) of the gapped Dirac point is occupied [see Fig. \ref{fig:HfCuGeAs}(a)], 
the large positive $\Omega_{xy}^{n}$ peaks [see the red curve in Fig. \ref{fig:HfCuGeAs}(d)] 
thus give rise to the dominant contribution to the total spin Berry curvature peaks 
at these $k$-points [see Fig. \ref{fig:HfCuGeAs}(e)].
Similar situations happen to $\Omega_{xz}^{n}$ and $\Omega_{zx}^{n}$ near the middle of the $M$-$\Gamma$ line
[see Figs. \ref{fig:HfCuGeAs}(f) and \ref{fig:HfCuGeAs}(g) as well as  
Figs. \ref{fig:HfCuGeAs}(h) and \ref{fig:HfCuGeAs}(i), respectively].
Figure \ref{fig:ZrCuGeAs} indicates that the calculated total and band-decomposed spin Berry curvatures in ZrCuGeAs
exhibit almost the same behavior as that of HfCuGeAs, although the peak magnitudes may differ. 

\begin{figure}[htbp] \centering 
\includegraphics[width=8.6cm]{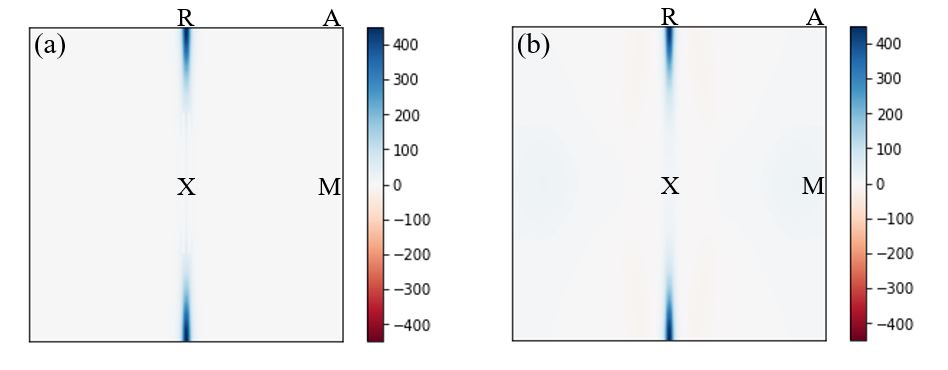}
\caption{Contour plot of (a) total and (b) Dirac line-node only contributions to spin Berry curvature $\Omega_{xy}({\bf k})$ of HfCuGeAs on the
\textit{X-R-A-M} plane [see Fig. 1(c)].}
\label{fig:HfCuGeAs-ContourPlot}
\end{figure}

In addition to the known resonance-like contributions from the slightly gapped Dirac points, we find that
the Dirac line-nodes can also make a significant contribution to the total spin Berry curvature and 
hence to SHC as well as SNC. For example, as mentioned above, there is a Dirac line-node along the $X$-$R$ line
slightly below the Fermi level in HfCuGeAs [see the red curve in Fig. \ref{fig:HfCuGeAs}(a)]. 
This line-node generates large spin Berry curvatures $\Omega_{xy}^{z}$, $\Omega_{xz}^{y}$, and $\Omega_{zx}^{y}$ 
along this $X$-$R$ line [see the red curve in Figs. \ref{fig:HfCuGeAs}(d), \ref{fig:HfCuGeAs}(f) and \ref{fig:HfCuGeAs}(h), respectively],
thus leading to the pronounced peaks in the corresponding total spin Berry curvatures on this line near the $R$ point
[see the blue curve in Figs. \ref{fig:HfCuGeAs}(e), \ref{fig:HfCuGeAs}(g) and \ref{fig:HfCuGeAs}(i), respectively].
The contour plots of the total and Dirac line-node spin Berry curvatures of $\Omega_{xy}({\bf k})$ 
on the $X$-$R$-$A$-$M$ plane in the BZ in HfCuGeAs 
are displayed in Figs. \ref{fig:HfCuGeAs-ContourPlot}(a) and \ref{fig:HfCuGeAs-ContourPlot}(b), respectively.
Figure \ref{fig:HfCuGeAs-ContourPlot} shows clearly that the total and Dirac line-node spin Berry 
curvatures $\Omega_{xy}({\bf k})$ of HfCuGeAs have large values only along the $X$-$R$ line.
Moreover, the total spin Berry curvature comes predominantly from the Dirac line-node spin Berry curvature. 
Therefore, this demonstrates that the occupied Dirac line-nodes near the Fermi level
can also contribute significantly to the large SHE and SNE in this Dirac line-node semimetal.

\section{CONCLUSIONS}

In conclusion, we have carried out a systematic \textit{ab initio} study on
the band structure topology, SHE and SNE in the quaternary arsenide XCuYAs (X=Zr, Hf; Y= Si, Ge) compounds 
based on relativistic density functional theory calculations.
Interestingly, we find that the four considered compounds are Dirac semimetals with the nonsymmorphic symmetry-protected Dirac
line nodes along the Brillouin zone boundary $A$-$M$ and $X$-$R$.
Furthermore, we also find that the intrinsic SHE and SNE in some of these considered compounds are large.
In particular, the calculated spin Hall conductivity (SHC) of HfCuGeAs is as large as -514 ($\hbar$/e)(S/cm).
The spin Nernst conductivity (SNC) of HfCuGeAs at room temperature is also large, being -0.73 ($\hbar$/e)(A/m-K).
Both the magnitude and sign of the SHC and SNC in these compounds are found to be tunable 
by varying either the applied electric field direction or spin current direction. The SHE and SNE
in these compounds can also be significantly enhanced by tuning the Fermi level via chemical doping or electric gating.
We have also calculated the band-decomposed and $k$-resolved spin Berry curvatures which reveal
that these large SHC and SNC as well as their notable tunabilities originate largely from
the presence of a large number of spin-orbit coupling-gapped Dirac points near the Fermi level as well as
the gapless Dirac line-nodes, which give rise to large spin Berry curvatures.
Our work thus shows that the four XCuYAs compounds not only provide a valuable platform
for exploring the interplay between SHE, SNE and band topology but also have promising applications
in spintronics and spin caloritronics.

\section*{ACKNOWLEDGMENTS}

The authors acknowledge the support from the Ministry of Science and Technology, National Center for Theoretical Sciences, 
and the Academia Sinica in Taiwan. G.-Y. Guo also thanks the support from the Far Eastern Y. Z. Hsu Science and Technology 
Memorial Foundation in Taiwan.

%

\end{document}